\documentclass[aip, jap, preprint]{revtex4-1}
\usepackage{amsmath,amssymb}
\usepackage{graphicx}

\begin{document}

% Use the \preprint command to place your local institutional report number 
% on the title page in preprint mode.
% Multiple \preprint commands are allowed.
%\preprint{}

\title{Surface Wave Transmission Line Theory for Single
and Many Wire Systems} %Title of paper

\author{Tobias Schaich}
\email[]{tcs49@cam.ac.uk}
\author{Daniel Molnar}
\author{Anas Al Rawi}
\altaffiliation[During time of research also with ]{BT Labs, Adastral Park, Martlesham
Heath, IP5 3RE, Ipswich, UK}%Lines break automatically or can be forced with \\
\author{Mike Payne}
\affiliation{ Cavendish Laboratory, University of Cambridge, CB3 0HE Cambridge, UK}%

\date{\today}% It is always \today, today,
             %  but any date may be explicitly specified
             
% repeat the \author .. \affiliation  etc. as needed
% \email, \thanks, \homepage, \altaffiliation all apply to the current author.
% Explanatory text should go in the []'s, 
% actual e-mail address or url should go in the {}'s for \email and \homepage.
% Please use the appropriate macro for the type of information

% \affiliation command applies to all authors since the last \affiliation command. 
% The \affiliation command should follow the other information.

\begin{abstract}
Examining cables using many conductor transmission
line theory has shed light on the modes supported by
various cable types. However, so far the theory disregards
the fundamental surface wave mode whose lateral confinement
increases with frequency and hence is expected to play an
important role in high frequency applications. To address this
issue, we propose an extension to the theory which incorporates
surface waves on uncoated, cylindrical wires. Crucially, this
requires new definitions of the per unit length transmission
line parameters which are derived using the single wire surface
wave solution. By closely examining a two wire and three wire
system, we show that these new parameters can predict surface
waves as well as modes found using conventional many conductor
transmission line theory. Furthermore, all calculated modes
are validated experimentally by diagonalization of a measured
channel transfer matrix. Additionally, the theoretically predicted
propagation constants for the modes are validated against full
numerical simulation for the two wire case and good agreement
is observed when proximity effects can be neglected.
\end{abstract}

% insert suggested PACS numbers in braces on next line

\maketitle %\maketitle must follow title, authors, abstract and \pacs

% If in two-column mode, this environment will change to single-column format so that long equations can be displayed. 
% Use only when necessary.
%\begin{widetext}
%$$\mbox{put long equation here}$$
%\end{widetext}

	\section{Introduction}
	The increasing need for high data rates necessary for many domestic applications has led to a push of the current copper access network namely in the form of Digital Subscriber Line (DSL) to higher frequencies. Current G.fast technology already uses frequencies up to 212~MHz with its successor G.mgfast pushing the maximum frequency even higher.\cite{ITU-T2019, Coomans2015, Oksman2019} Moreover, recent experimental results using terahertz signals indicated that terabit DSL may be possible.\cite{Phys2020} This push to higher frequencies has been facilitated by advances in digital signal processing, more specifically in the form of vectoring and treating cables as multiple-input multiple-output (MIMO) channels.\cite{Ginis2002,Lee2007} Even so, these technological advances require understanding the full potential of the mixed-mode copper channel beyond the state of the art instrumentation and accurate channel models at low to high frequencies in order to predict direct transmission and crosstalk levels of the MIMO channel. 
	
	A successful simulation method for cables is based on many conductor transmission line (MTL) theory. Shielded and unshielded twisted pairs as well as more complicated binder cables have been examined with this method.\cite{Paul1979, Brandao2003a, Strobel2013, Lee2007} It combines a simple and scalable approach which allows the treatment of both uniform and non-uniform lines making it a versatile tool for modelling many types of cable.\cite{Paul1994}
	
	However, as frequency increases modes which are disregarded by MTL may start propagating. In particular, surface waves which are fundamental modes of single wires have been disregarded as their lateral size is extremely large for low frequencies.\cite{Sommerfeld1899} But as frequencies reach the MHz and GHz scale, these modes can start to play a role in the transmission characteristics of binder cables. Hence, in this work we attempt to extend the framework of MTL to include surface waves.
	
	 To this end, Section \ref{sec:TLModel} shows that single wire surface waves may be treated within scalar transmission line theory. In Section \ref{sec:MTLModel}, we extend our results to the matrix MTL theory. Section \ref{sec:Geometries} applies the theory to a uniform 2-wire and a 3-wire system and includes validation against experimental and numerical results. We conclude with a quick summary and outline future avenues of research.	
	 
	 \section{Transmission Line Model for Sommerfeld Surface Wave} \label{sec:TLModel}
	This section formulates a transmission line equivalent circuit with distributed circuit elements to represent the Sommerfeld surface wave which can exist on a single uncoated conductor.
	
	\subsection{The Sommerfeld Surface Wave}
	An infinite, uncoated wire in vacuum with radius $a$ and conductivity $\sigma$ supports a surface wave with electric and magnetic fields $\vec{E}$ and $\vec{H}$.\cite{Sommerfeld1899, Goubau1950} Harmonic time dependence $e^{i\omega t}$ with angular frequency $\omega$ and time $t$ and a cylindrical coordinate system $(r,\phi,z)$ with origin at the centre of wire is used throughout. Furthermore, propagation of the form $e^{-i\beta z}$ with propagation constant $\beta$ is assumed but omitted for brevity. Then, the non-zero field components of $\vec{E}$ and $\vec{H}$ are:
	\begin{subequations}
	\begin{equation}
		E_z=\begin{cases}
			A J_0(\gamma_c r) \quad r\leq a,\\
			B H_0(\gamma r) \quad r\geq a,
		\end{cases} \label{eq:Sommerfeld_z}
	\end{equation}
	\begin{equation}
		E_r=\begin{cases}
			A \frac{i \beta}{\gamma_c} J_1(\gamma_c r) \quad r\leq a, \\
			B \frac{i \beta}{\gamma} H_1(\gamma r) \quad r > a,
		\end{cases}\label{eq:Sommerfeld_r}
    \end{equation}
	\begin{equation}
		H_\phi=\begin{cases}
			A \frac{i \omega \epsilon_c}{\gamma_c} J_1(\gamma_c r) \quad r\leq a, \\
			B \frac{i \omega \epsilon_0}{\gamma} H_1(\gamma r) \quad r\geq a,
		\end{cases}\label{eq:Sommerfeld_phi}
	\end{equation}
	\end{subequations}
	where $i=\sqrt{-1}$ and $J_0$, $J_1$ are Bessel functions of the first kind and order 0 and 1, respectively.\cite{Goubau1950, Orfanidis2004} Similarly, $H_0$ and $H_1$ are Hankel functions of the first kind and order 0 and 1. In the derivation, the wire is treated as a lossy dielectric with permittivity $\epsilon_c=\epsilon_0-i\frac{\sigma}{\omega}$ and vacuum permittivity $\epsilon_0$. The permeability of vacuum is $\mu_0$. $A$ and $B$ are amplitudes with units of V/m. Finally, $\gamma=\sqrt{\omega^2\epsilon_0 \mu_0 - \beta^2} $ and $\gamma_c=\sqrt{\omega^2\epsilon_c \mu_0 - \beta^2}$ are the transverse decay constants in the surrounding vacuum and the conductor, respectively. The tangential electric and magnetic fields are continuous at the boundary $r=a$ which relates the amplitudes $A$ and $B$. Thus, we can formulate a transcendental equation for the propagation constant:
	\begin{equation}
		\beta^2=\omega^2\epsilon_0 \mu_0 - \frac{\epsilon_0 \gamma_c \gamma J_0(\gamma_c a) H_1(\gamma a) }{\epsilon_c J_1(\gamma_c a) H_0(\gamma a)}.  \label{eq:characteristic}
	\end{equation}
	The propagation constant for the Sommerfeld wave is complex. For waves travelling in the positive $z$-direction, $\beta = \beta_r - i \beta_i$ with real, positive values for $\beta_r$ and $\beta_i$, which are known as the phase constant and attenuation constant, respectively. Both vary with the frequency of the surface wave. 

	\subsection{Definition of Voltage and Current}
	
	Now that the fields have been established, the first step towards a transmission line model is the definition of a voltage and current. We shall start by considering the current. Using the microscopic version of Maxwell's equations, Ampere's law gives the current density $\vec{\mathcal{J}}$ as:
	\begin{equation}
		\vec{\mathcal{J}}(r,\phi,z)=\nabla \times \vec{H} - i \omega \epsilon_0 \vec{E}=\begin{cases}
			\sigma \vec{E} \quad r\leq a, \\
			0 \quad r>a,
		\end{cases} \label{eq:current-density}
	\end{equation}
	 where for the last equality, we have used the fields for the Sommerfeld wave \eqref{eq:Sommerfeld_z}-\eqref{eq:Sommerfeld_phi}. Thus, $\vec{\mathcal{J}}$ has components $\mathcal{J}_z$ and $\mathcal{J}_r$. The current is now defined as the axial current density integrated over a surface $X$ which encompasses the entire conductor cross-section and whose normal vector points in the positive $z$-direction. In particular, this gives (omitting $e^{-i\beta z}$)
	\begin{equation}
		I(z)=\int_X \mathcal{J}_z r dr d\phi = A  \frac{2\pi a \sigma}{\gamma_c} J_1 (\gamma_c a). \label{eq:current}
	\end{equation}
	We note that the first equality is general in nature and the second equality uses \eqref{eq:current-density}. Another useful expression is found by using continuity of the tangential magnetic field at $r=a$ allowing to rewrite the current as:
	\begin{equation}
		I(z)=2 \pi a B \frac{\sigma \epsilon_0}{\epsilon_c} \frac{H_1(\gamma a)}{\gamma}. \label{eq:current-simple}
	\end{equation}
	
	The voltage $V$ is defined, in analogy to the definition for transverse electromagnetic (TEM) waves, as the line integral over the electric field along any contour $C$ lying  in the transverse plane going from $r=a$ to $r=\infty$,\cite{Paul1994} i.e.
	\begin{equation}
		V(z)=\int_C \vec{E}(r,\phi,z) \cdot d\vec{l} = B\frac{i\beta}{\gamma^2}H_0(\gamma a). \label{eq:voltage}
	\end{equation}
	Again the first equality is general while the second uses \eqref{eq:Sommerfeld_r}. This definition is unique for the transverse magnetic surface wave because it does not have any azimuthal electric field components which means only the radial dependence enters the integral. Finally, we remark that this voltage definition places the reference or ground infinitely far away from the wire. We reiterate that this definition of voltage is equally applicable for TEM waves.
	
	Now that current and voltage have been defined, we will derive equations, known as telegrapher's equations, which relate them on a transmission line.
	
	\subsection{Derivation of Telegrapher's Equations} \label{sec:TL_Eq}
	\begin{figure}
		\centering
		\includegraphics[width=0.45\textwidth]{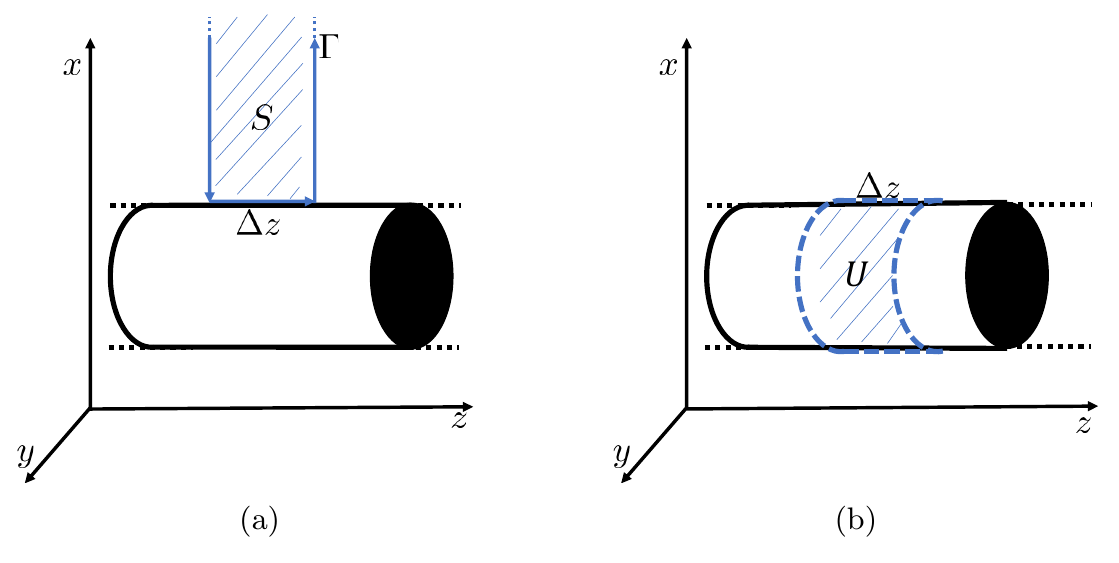}
		\caption{(a) Surface $S$ with bounding contour $\Gamma$ for the derivation of first telegrapher's equation. (b) Volume $U$ for the derivation of second telegrapher's equation.}
		\label{fig:TL-geometries}
	\end{figure}
	For the first transmission line equation, consider the surface $S$ shown in Fig. \ref{fig:TL-geometries}(a). Integrating Faraday's law over this surface gives
	\begin{equation}
		\int_S (\nabla \times \vec{E})\cdot d\vec{S} = -i \omega \mu_0 \int_S \vec{H} \cdot d\vec{S}, \label{eq:Faraday-integrated}
	\end{equation}
	where the infinitesimal surface element $d\vec{S}$ points in the azimuthal direction. For the left hand side of \eqref{eq:Faraday-integrated}, we invoke Stokes' theorem and turn it into a line integral over the boundary of $S$ called $\Gamma$ in Fig. \ref{fig:TL-geometries}(a). Moreover, as the fields are necessarily zero at infinity in order for the system to have finite energy, the integral may be written as:
	\begin{eqnarray}
		\int_S (\nabla \times \vec{E}) \cdot d\vec{S}=\int_{a}^{\infty}\Big(E_r(z+\Delta z) - E_r(z)\Big) dr \nonumber \\
		+ \int_{z}^{z+\Delta z} E_{z}(z') dz'. \label{eq:faraday-curl}
	\end{eqnarray}
	The first two terms can be identified as voltages according to our definition in \eqref{eq:voltage}. 
	
	Ultimately we will take the limit $\Delta z \rightarrow 0$ to define per unit length circuit elements. So, it is useful to now introduce a few definitions. Clearly, the integral on the right hand side of \eqref{eq:Faraday-integrated} describes the magnetic flux through the surface $S$. This can be related to the current in \eqref{eq:current-simple} which allows the definition of a per unit length self-inductance $L$ as:
	\begin{equation}
		L=\lim\limits_{\Delta z \rightarrow 0} \mu_0 \frac{\int_S \vec{H} \cdot d\vec{S}}{\Delta z I(z)}=\mu_0\frac{ i\omega \epsilon_c H_0(\gamma a)}{2\pi \gamma a \sigma H_1 (\gamma a)}. \label{eq:inductance}
	\end{equation}
	The advantage of this formulation is that $L$ is no longer dependent on $z$ as the propagation factors cancel.

	 Another useful definition concerns the integral containing the longitudinal electric field in \eqref{eq:faraday-curl}. This can be related to the current in \eqref{eq:current} to create a per unit length conductor impedance $Z_R$ which is defined as:
	\begin{equation}
		Z_R=\lim\limits_{\Delta z \rightarrow 0}\frac{\int_{z}^{z+\Delta z}E_z(z') dz'}{\Delta z I(z)}= \frac{1}{2\pi a \sigma} \frac{\gamma_c J_0(\gamma_c a)}{J_1(\gamma_c a)}. \label{eq:Impedance}
	\end{equation}
	With these two definitions, we may divide \eqref{eq:Faraday-integrated} by $\Delta z$ and take the limit $\Delta z \rightarrow 0$. Thus, we arrive at the first telegrapher's equation:
	\begin{equation}
		\frac{\partial V(z)}{\partial z} = -Z_R I(z) - i \omega L  I(z). \label{eq:TL1}
	\end{equation}
	For clarity dependency on the $z$-coordinate was made explicit.
	
	To derive the next transmission line equation, we invoke the continuity equation 
	\begin{equation}
		\nabla \cdot \vec{\mathcal{J}}=-i \omega \rho, \label{eq:continuity-diff}
	\end{equation}
	where $\rho$ is the charge density. Integrating over the volume $U$ shown in Fig. \ref{fig:TL-geometries}(b) and using the divergence theorem, the left hand side separates into an integral over the cylinder caps coinciding with the cross-section $X$ and over the mantle $M$. We define $U$ such that the mantle is just outside the conductor. Given \eqref{eq:current-density} the mantle term then vanishes, leaving
	\begin{equation}
		\int_{V}\nabla \cdot \vec{\mathcal{J}} dV = \int_X \Big(\mathcal{J}_z(z+\Delta z).
	\end{equation}
	With the definitions in \eqref{eq:current} and the current density of the Sommerfeld wave in \eqref{eq:current-density}, the integrated version of \eqref{eq:continuity-diff} becomes:
		\begin{equation}
			I(z+\Delta z)-I(z)=-i \omega \int_U \rho dU. \label{eq:continuity-int}
		\end{equation}
	The total charge $Q=\int \rho dU$ enclosed in the volume $U$ is found on the right hand side of \eqref{eq:continuity-int}. This charge may be related to the voltage to form a capacitance. As before, we will be considering the limit $\Delta z\rightarrow 0$ and so we may define a per unit length capacitance as:
	\begin{equation}
		C=\lim\limits_{\Delta z \rightarrow 0} \frac{Q}{\Delta z V(z)}.
	\end{equation}
	To calculate the enclosed per unit length charge, we use Gauss' law and the divergence theorem to split the volume integral into a surface integral over the mantle and the caps. After some simplifications, we find
	\begin{equation}
		\lim \limits_{\Delta z \rightarrow 0}\frac{Q}{\Delta z}=2\pi  a \epsilon_0 B\frac{i\beta}{\gamma}H_1(\gamma a)(1-\frac{\epsilon_0}{\epsilon_c}). \label{eq:charge}
	\end{equation}
	This result is proportional to the discontinuity of the radial electric field across the boundary $r=a$ and can be associated with a surface charge on the conductor. Using \eqref{eq:charge}, the per unit length capacitance may be expressed as:
	\begin{equation}
		C= \epsilon_0 \frac{2 \pi \gamma a \sigma H_1(\gamma a)}{i \omega \epsilon_c H_0(\gamma a)}  ,
	\end{equation} 
	where in passing we note that $C=\mu_0 \epsilon_0 L^{-1}$. Returning to \eqref{eq:continuity-int}, we divide by $\Delta z$ and take the limit $\Delta z\rightarrow 0$. Using the derived per unit length capacitance, we get the second telegrapher's equation which is of the form 
	\begin{equation}
		\frac{\partial I(z)}{\partial z}= - i \omega C V(z). \label{eq:TL2}
	\end{equation} 
	
	Note that the frequency-domain results may be converted to time-domain using an inverse Fourier transform assuming the frequency dependent amplitudes are known.\cite{Wedepohl1969, Gordon1992}
	
	Furthermore, we remark that the derived per unit length inductance and capacitance are frequency dependent. In contrast, (lossless) TEM or quasi-TEM waves have frequency-independent and real per unit length capacitance and inductance. In fact, the derived per unit length parameters for the surface wave diverge as the frequency tends to zero. This is due to the increasing delocalization of surface waves with decreasing frequency. At zero frequency, no surface wave solution exists.
	
	 Finally, we want to emphasise that the calculated per unit length inductance, capacitance and impedance are all generally complex valued. The complex behaviour of these quantities can in part be understood by decomposing them into real and imaginary parts. The imaginary part of the impedance $Z_R$ has been shown to be linked to an internal inductance inside the wire.\cite{Paul1994} The more unusual complex $C$ can be separated into a capacitive and conductive part. Assuming $i\omega C=i\omega C' + G$ with a real-valued per unit length capacitance $C'$ and conductance $G$, we solve for $G$ to find 
	 \begin{equation}
	 	G=-\omega \text{Im}(C)=2\pi a \text{Re}\Big(\frac{\mathcal{J}_r(r=a)}{V}\Big), 
	 \end{equation}
 	 where Im and Re indicate the imagary and real part, respectively.
	 This shows that the conductance is related to the in-phase component of the radial current density with the voltage. As $LC=\mu_0 \epsilon_0$ for the complex $L$ and $C$, this implies further that there is a resistance $R_L$ of opposite sign to the conductance $G$ associated with the complex inductance $L$. In all cases studied, we found a positive conductance $G$ and negative resistance $R_L$. However, a negative $R_L$ does not lead to unphysical behaviour such as gain from a passive transmission line as will be shown in the following section.
	
	\subsection{Solution to the Telegrapher's equation}
	At non-zero frequencies, it is possible to utilise results based on the standard solution of the telegrapher's equations \eqref{eq:TL1} and \eqref{eq:TL2} in frequency domain.\cite{Paul1994} In particular, we may infer that voltage and current waves exist and propagate in space as $e^{-i\beta_{TL}z}$ with
	\begin{equation}
		\beta_{TL}^2=- (Z_R+i\omega L)(i\omega C). \label{eq:characteristic-TL}
	\end{equation}
	Arguably, $\beta_{TL}$ will coincide with the value of the analytic propagation constant of the Sommerfeld wave $\beta$. Indeed equations \eqref{eq:characteristic-TL} and \eqref{eq:characteristic} are equal requiring $\beta_{TL}=\beta$. If $Z_R$ were zero, $LC=\mu_0\epsilon_0$ would imply lossless propagation at the speed of light in vacuum. Consequently, losses and a sub light speed phase velocity result from the wire impedance $Z_R$ namely through the term $-i\omega C Z_R$. Thus, the negative resistance $R_L$ associated with the complex inductance $L$ does not affect the propagation losses. 
		
	Additionally, we may define a characteristic impedance of the surface wave $Z_0$ which relates current and voltage waves travelling in the same direction. It can be calculated as:
	\begin{equation}
		Z_0=\sqrt{\frac{Z_R+i\omega L}{i \omega C}}
	\end{equation}
	and is generally complex and frequency dependent. The characteristic impedance in a quasi-TEM transmission line is a useful quantity because reflections along the transmission line can be linked to discontinuities in the characteristic impedance. Moreover, impedance matching can be used to mitigate such reflections. For the case of surface waves, impedance matching is more difficult as the ground is at infinity. Hence, matching to a standard transmission line with ground such as a coplanar or a microstrip line cannot be reduced to an impedance discontinuity calculation. Furthermore, for most geometries surface waves are only weakly bound to the wire leading to a non-negligible influence of radiation modes at discontinuities. In spite of this, the characteristic impedance of a surface wave may still hold value as a heuristic quantity for the design of surface wave launchers and interconnects. 
	
	Finally, consider the power  $P_T$ transmitted through a cross-section outside the wire in the $z$-direction. Using Poynting's theorem and the fields of the Sommerfeld wave, the power is calculated as:\cite{Orfanidis2004}
	\begin{equation}
		P_T=\frac{1}{2}|B|^2\frac{2\pi a \epsilon_0}{|\gamma|^2} \frac{\omega \beta_r \text{Im}(\gamma   H_0(\gamma a )H_1(\gamma a)^*)}{\text{Im}(\gamma ^2)}, \label{eq:PT}
	\end{equation}
		where $^*$ denotes complex conjugation. Let us compare this to the power delivered using transmission line theory $P_{TL}=\frac{1}{2}\text{Re}(VI^*)$. Using \eqref{eq:current} and \eqref{eq:voltage} this power is
		\begin{equation}
			P_{TL}=\frac{1}{2}|B|^2\frac{2\pi a \epsilon_0}{|\gamma|^2} \text{Re}\Big(\frac{i \beta \sigma}{\epsilon_c^*} \frac{\gamma H_0(\gamma a) H_1(\gamma a)^*}{\gamma^2}\Big). \label{eq:PTL}
		\end{equation} 
	$P_{TL}$ is not generally equal to $P_T$ but it approximates $P_T$ closely in the common case where the wire is a good conductor and the Sommerfeld wave travels with small loss ($\beta_r \gg \beta_i$) close to the speed of light (see Appendix \ref{app:Pt=Ptl}). This further validates our approach and supports the argument that surface waves can be treated within transmission line theory.
	
		\section{Many-Conductor Transmission Line Model} \label{sec:MTLModel}
	
	So far, we have shown that a transmission line equivalent exists for the surface wave on a single wire which reproduces the analytic propagation constant. A natural extension of this problem is to consider multiple identical wires each able of supporting a surface wave by itself. Thus, the single wire scalar quantities $C$, $L$ and $Z_R$ must be generalised to matrix quantities $\mathbf{C}$, $\mathbf{L}$ and $\mathbf{Z_R}$ which can be used in many conductor transmission line theory.\cite{Pipes1937, Paul1994}
	
	Consequently, we consider an $N$ wire system, where all wires have conductivity $\sigma$ and a radius $a$. Furthermore, let $d_{mn}$ be the distance between the centres of wire $m$ and wire $n$. Let $I_n$ and $V_n$ be the current and voltage on the $n$-th wire as defined generally in \eqref{eq:current} and \eqref{eq:voltage} (first equality). Then, we define a current and voltage vector as:
	\begin{align}
	\underline{I}=\begin{pmatrix}
			I_1 \\ \vdots \\ I_N 
		\end{pmatrix}, &&
	\underline{V}=\begin{pmatrix}
			V_1 \\ \vdots \\V_N
		\end{pmatrix} \label{eq:I-V-vectors},
	\end{align}  
	where we use an underscore notation to distinguish them from 3-dimensional spatial vectors. Let all wires be sufficiently separated ($d_{mn}\gg a$) such that proximity effects can be neglected. Then, the electromagnetic field outside the wires can be approximated by a superposition of surface waves. As the current density of a single-wire surface wave \eqref{eq:current-density} is zero outside its wire, we approximate that the current on any wire is simply due to the surface wave propagating on this wire. Hence, we may use the currents of a single-wire surface wave from \eqref{eq:current} or \eqref{eq:current-simple}. In contrast, the voltage of a wire will depend on the surface waves on itself and all other wires. So, the second equality in \eqref{eq:voltage} does not hold in general. Voltages and currents are related by many-conductor transmission line (MTL) telegrapher's equations, which are derived in the following paragraphs. 
	 	
	The derivation of the first telegrapher equation is analogous to the single wire case in section \ref{sec:TL_Eq}. We let $S_m$ be a surface touching conductor $m$ as in Fig. \ref{fig:TL-geometries}(a) and integrate Faraday's law over this surface to get
	\begin{equation}
		\int_{S_m} \big(\nabla \times \vec{E_T}\big)\cdot d\vec{S} = -i \omega \mu_0  \int_{S_m} \vec{H}_T \cdot d\vec{S}, \label{eq:Faraday-Int-MTL}
	\end{equation}
	 where $\vec{E_T}$ and $\vec{H_T}$ are the total electric and magnetic fields which consist of surface wave contributions from all wires. 
	 
	 The right hand side of \eqref{eq:Faraday-Int-MTL} contains the magnetic flux through $S_m$. As in the single wire case, the surface wave on conductor $m$ will contribute $L I_m$ to this flux where $L$ is the self-inductance derived in \eqref{eq:inductance}. However, the presence of other surface waves can now contribute to the overall magnetic flux through the surface $S_m$ as well. Let $L_{mn} I_n$ be the additional magnetic flux caused by the surface wave on wire $n$ with current $I_n$. The mutual inductance $L_{mn}$ is defined in analogy to \eqref{eq:Faraday-integrated} as:
	\begin{equation}
		L_{mn}=\lim\limits_{\Delta z \rightarrow 0} \mu_0 \frac{\int_{S_m}\vec{H}_n \cdot d\vec{S}}{\Delta z I_n(z)}=\mu_0 \frac{H_0(\gamma d_{mn})}{2\pi a \gamma H_1(\gamma a)},
	\end{equation}
	where $\vec{H}_n$ denotes the magnetic field of the surface wave on wire $n$. For the last equation we have used \eqref{eq:Sommerfeld_phi} and assumed the finite size of the wires is negligible as $d_{mn}\gg a$. In that case the value of the integral is also identical for any surface $S_m$ going from the conductor $m$ to infinity which follows directly from Gauss' law for magnetic fields. So, the total magnetic flux through surface $S_m$ is given by 
	\begin{equation}
		\mu_0\int_{S_m}\vec{H}_T \cdot d\vec{S}=\sum_{n=1}^{N}L_{mn} I_n,
	\end{equation}
	where $L_{mm}=L$ from \eqref{eq:inductance}. 
	
	Using Stokes' theorem on the left hand side of \eqref{eq:Faraday-Int-MTL}, the integral is transformed into two radial integrals and a longitudinal integral as in \eqref{eq:faraday-curl}. With the definition of voltage in \eqref{eq:voltage}, the radial integrals give the voltage difference along the line $V_m(z+\Delta z) - V_m(z)$. The longitudinal integral can be associated with a per unit length wire impedance as in \eqref{eq:Impedance}. Naively, a component surface waves from wire $n$ would contribute a factor $Z_{mn}$ to the impedance as the integral involves the total electric field. However, longitudinal electric fields from all surface waves on the wire imply that its current depends on all component surface waves. This contradicts the previous assumption that the current of a wire only depends on the surface wave which propagates on said wire. Hence, we postulate that the longitudinal field of any surface waves bound to wire $n$ must be zero on wire $m$ if $m\neq n$. Thus, the impedance $Z_{mn}$ is
	\begin{align}
		Z_{mn}=\begin{cases}
			0 \quad & \text{if $m\neq n$}\\
			Z_R \quad & \text{if $m=n$}
		\end{cases},
	\end{align}
	with $Z_R$ given in \eqref{eq:Impedance}.
	 Interpreting $Z_{mn}$ and $L_{mn}$ as matrix elements of $\mathbf{Z_R}$ and $\mathbf{L}$ respectively, the telegraphers equation in matrix form can be written concisely as:
	\begin{equation}
		\frac{\partial \underline{V}}{\partial z}=\big(-\mathbf{Z_R} - i \omega \mathbf{L}\big) \underline{I}.
		\label{eq:MTL-1}
	\end{equation}
	
	For the second telegrapher's equation, we consider a volume $U_m$ around wire $m$ of infinitesimal length $\Delta z$ in analogy to the single wire case in Fig. \ref{fig:TL-geometries}(b). Integrating the continuity equation over this volume will result in the following equation:
	\begin{equation}
		I_m(z+\Delta z)-I_m(z)=-i\omega Q_m , \label{eq:continuity-int-MTL}
	\end{equation}
	where $Q_m$ is the charge on conductor $m$ inside the volume $U_m$. The similarity to \eqref{eq:continuity-int} is apparent. Our aim is to relate current with voltages. So, we assume there are per unit length capacities $C_{mn}$ which connect the charge on conductor $m$ to the voltage on conductor $n$ via
	\begin{equation}
		\lim\limits_{\Delta z \rightarrow 0} \frac{Q_m}{\Delta z}=\sum_{n=1}^N C_{mn} V_n. \label{eq:charge-MTL}
	\end{equation} 
	Under the assumption that the fields are a superposition of surface waves, the charge $Q_m$ only depends on the surface wave on $m$ and can be calculated from Gauss' law and the Sommerfeld solution. Then, the elements $C_{mn}$ are found by forcing all voltages except $V_n$ to zero and solving \eqref{eq:charge-MTL}. An alternative calculation of the elements $C_{mn}$ using coefficients of potential is outlined in Appendix \ref{app:CfromP}. There, we also show that for good conductors ($\frac{\sigma}{\omega}\gg\epsilon_0$) we may approximate $\mathbf{C}\approx \mu_0 \epsilon_0 \mathbf{L}^{-1}$ where $\mathbf{C}$ has elements $C_{mn}$. Returning to \eqref{eq:continuity-int-MTL}, we divide by $\Delta z$ and take the limit $\Delta z \rightarrow 0$ which permits use of the sum in \eqref{eq:charge-MTL}. The result is the second telegrapher's equation for any wire $m$. Writing the equation in matrix form gives
	\begin{equation}
		\frac{\partial \underline{I}}{\partial z}=-i\omega \mathbf{C} \underline{V}. \label{eq:MTL-2}
	\end{equation}
	
	The equations \eqref{eq:MTL-1} and \eqref{eq:MTL-2} form the basis of MTL theory which has been discussed in great detail in many books and papers of which a non-exhaustive list is referenced here.\cite{Pipes1937, Pipes1941, Wedepohl1963, Wedepohl1969, Leviatan1982, Harrington1984, Gordon1992, BrandaoFaria1993, Paul1994, BrandaoFaria2014} Most works on the topic calculate the per unit length transmission line parameter matrices $\mathbf{L}$, $\mathbf{C}$ and $\mathbf{Z_R}$ from a quasi-static approach which can be understood as the first order term in a expansion with frequency also known as the quasi-TEM field.\cite{Lindell1981} Therein lies the fundamental difference of the derived quantities in this work. For surface waves, there is no first order or quasi-TEM field as a static surface wave cannot exist.\cite{Sommerfeld1899} Therefore, this theory is fundamentally frequency dependent and distinguished from previous studies. However, many methods used in prior works such as the cascading of uniform segments to simulate a non-uniform cable are applicable.\cite{Lee2007} Nonetheless, in the next section we focus on the eigenmodes of some simple, uniform systems and develop our models connection to previous MTL studies. 
	
	\section{Eigenmodes of Select Geometries} \label{sec:Geometries}
	
	In this section, we investigate the eigenmodes of a selection of uniform geometries theoretically and experimentally. For the theoretical investigation, the modal voltages are found as the eigenvectors of the matrix $(\mathbf{Z_R}+i\omega \mathbf{L})(i \omega \mathbf{C})$ and the eigenvalues describe the negative square of the propagation constant for this mode i.e. $-\beta^2$ (Refs. \onlinecite{BrandaoFaria1993, Paul1994, BrandaoFaria2014}).  Modal currents given by the eigenvectors of $(i \omega \mathbf{C})(\mathbf{Z_R}+i\omega \mathbf{L})$ are nearly identical to the voltage eigenvectors as $\mathbf{Z_R}$ is diagonal and $\mathbf{L}$ and $\mathbf{C}$ commute implied by $\mathbf{C}\approx \mu_0 \epsilon_0 \mathbf{L}^{-1}$ (Refs. \onlinecite{Paul1979, Gentili1995}). Hence, no distinction between voltage and current eigenvectors is made. 
	
\begin{figure}[t]
		\centering
		\includegraphics[width=0.45\textwidth]{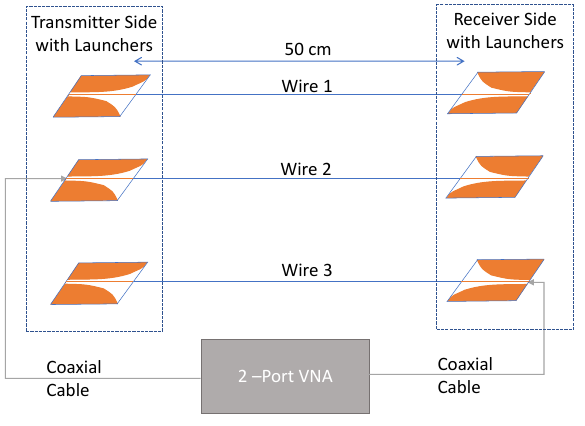}
		\caption{Schematic three wire setup to determine the transfer matrix $\mathbf{H}$ shown here in a configuration to measure $H_{32}$. In the two wire setup, Wire 3 and its launchers are removed.}
		\label{fig:Setup}
	\end{figure}

	To experimentally determine the eigenmodes we rely on a concept from multiple-input multiple-output (MIMO).\cite{Lee2007} In our setup, copper wires with 0.5 mm diameter are soldered to a pair of surface wave launcher which allow a smooth impedance transformation to a 50~$\Omega$ SMA connector. Each launcher pair is slotted into a plastic jig such that the launchers face each other and the connecting wire is as straight as possible. The jig allows to laterally place the launchers a certain distance apart and elevates the wire above the lab bench. Wire lengths of 50 cm were chosen to avoid sag which could lead to radiation and mode coupling. Images of the setup and launchers are provided in the Supplementary Material. With the launchers in place, a transmitting and receiving side is chosen arbitrarily. Sequentially, each combination of transmitting and receiving launcher is connected to a 2-port Vector Network Analyser with coaxial cables and its scattering parameters are measured \cite{Schaich2021}. Unconnected ports are terminated with 50 $\Omega$ loads. A schematic of the measurement setup for three wires is shown in Fig. \ref{fig:Setup}. 
	
	The scattering parameter related to an excitation of wire $n$ and a measurement of wire $m$ will be denoted $H_{mn}$. For a system of $N$ wires, this results in an $N\times N$ channel matrix $\mathbf{H}$. Its eigenvectors describe the eigenmodes of the system as they, by definition, traverse the system unaltered except for potential losses and phase shifts. Thus, the normalised eigenvectors of $\mathbf{H}$ should coincide with the normalised voltage eigenvectors.
	
	\subsection{Two Wires}

	The simplest non-trivial case consists of two identical wires which are separated by a distance $d$. Due to the symmetry of the problem, the matrices $\mathbf{L}$, $\mathbf{Z_R}$ and $\mathbf{C}$ are all symmetric with identical elements along the diagonal. Thus, they all commute and can be simultaneously diagonalised by a matrix $\mathbf{T}$ with
	\begin{equation}
	 \mathbf{T}=\mathbf{T}^{-1}=\frac{1}{\sqrt{2}} \begin{pmatrix} 1 & 1\\1	& -1\end{pmatrix}, 
	\end{equation}
	whose column vectors coincide with the voltage and current eigenmodes of the system. Focussing on the first eigenvector, it corresponds to in phase excitation with equal currents and voltages being applied to both conductors at one end. Its form is closely related to the common mode in grounded circuits. However, there is no ground in this setup which could carry a return current. Rather, this mode behaves analogous to the single wire surface wave. Therefore, we identify it as a multi-conductor surface wave mode. In contrast, in the second eigenvector both wires are excited $\pi$ out of phase meaning they  always carry an equal but opposite voltage and current. This mode is known in the literature as differential mode.\cite{Bockelman1995, Lee2007}
	
	This distinction into surface wave mode and differential mode will now be further investigated. To that end, let us calculate the diagonalised per unit length parameters. Due to the symmetry property discussed above, unambiguous definition of modal inductance and capacitance are possible \cite{Gentili1995}. The surface wave inductance $L_{SW}$ and a differential mode inductance $L_{DM}$ are given via
	\begin{equation}
	\mathbf{T}^{-1}\mathbf{L}\mathbf{T}=
		\begin{pmatrix}
			L_{SW}  & 0 \\
			0 & L_{DM}
		\end{pmatrix}.
	\end{equation}
	Assuming a good conductor ($\frac{\sigma}{\omega} \gg \epsilon_0$), the inductances are
	\begin{gather}
		L_{SW}=\mu_0\frac{H_0(\gamma a) + H_0(\gamma d)}{2\pi \gamma a H_1(\gamma a)} \label{eq:inductance-SW}, \\
		L_{DM}=\mu_0 \frac{H_0(\gamma a) - H_0(\gamma d)}{2\pi \gamma a H_1(\gamma a)}. \label{eq:inductance-DM}
	\end{gather}
	The respective capacitances are related to the inductances via $\mathbf{C}=\mu_0 \epsilon_0 \mathbf{L}^{-1}$ which also holds for the diagonalised version of $\mathbf{L}$. The resistance matrix $\mathbf{Z_R}$ remains unchanged after diagonalization. 
	
	Let us focus on the differential mode inductance and consider the limit $|\gamma a| \ll 1$ signifying a low loss surface wave propagating near the speed of light. Then, we may use asymptotic expressions for the Hankel function in \eqref{eq:inductance-DM} which simplifies the inductance to \cite{Abramowitz1972}
	\begin{equation}
		L_{DM}\approx \frac{\mu_0}{2\pi} \log{\Big(\frac{d}{a}\Big)}.
	\end{equation}
	This is exactly half the inductance of a two wire line with one conductor at ground found in literature. \cite{Paul1994} The factor of $1/2$ is due to the definition of the ground which in our theory sits at infinity or, in the case of the differential mode, may also be placed in the centre between the wires. Also, the frequency dependence has dropped out of the inductance. Crucially, this result reinforces the assertion that the mode which we have identified as a differential mode is indeed the mode which has been discussed previously in literature. Consequently, the presented theory fits in well with previous studies using the MTL model and will reproduce similar results in cases where the contribution of the surface wave can be neglected. In fact, as surface waves can easily radiate when perturbed, their contribution may have been interpreted as radiation in previous studies.\cite{Chang1976, Nakamura1994, Schaich2020b}
	
	\begin{figure}
		\centering 
		\includegraphics[width=.45\textwidth]{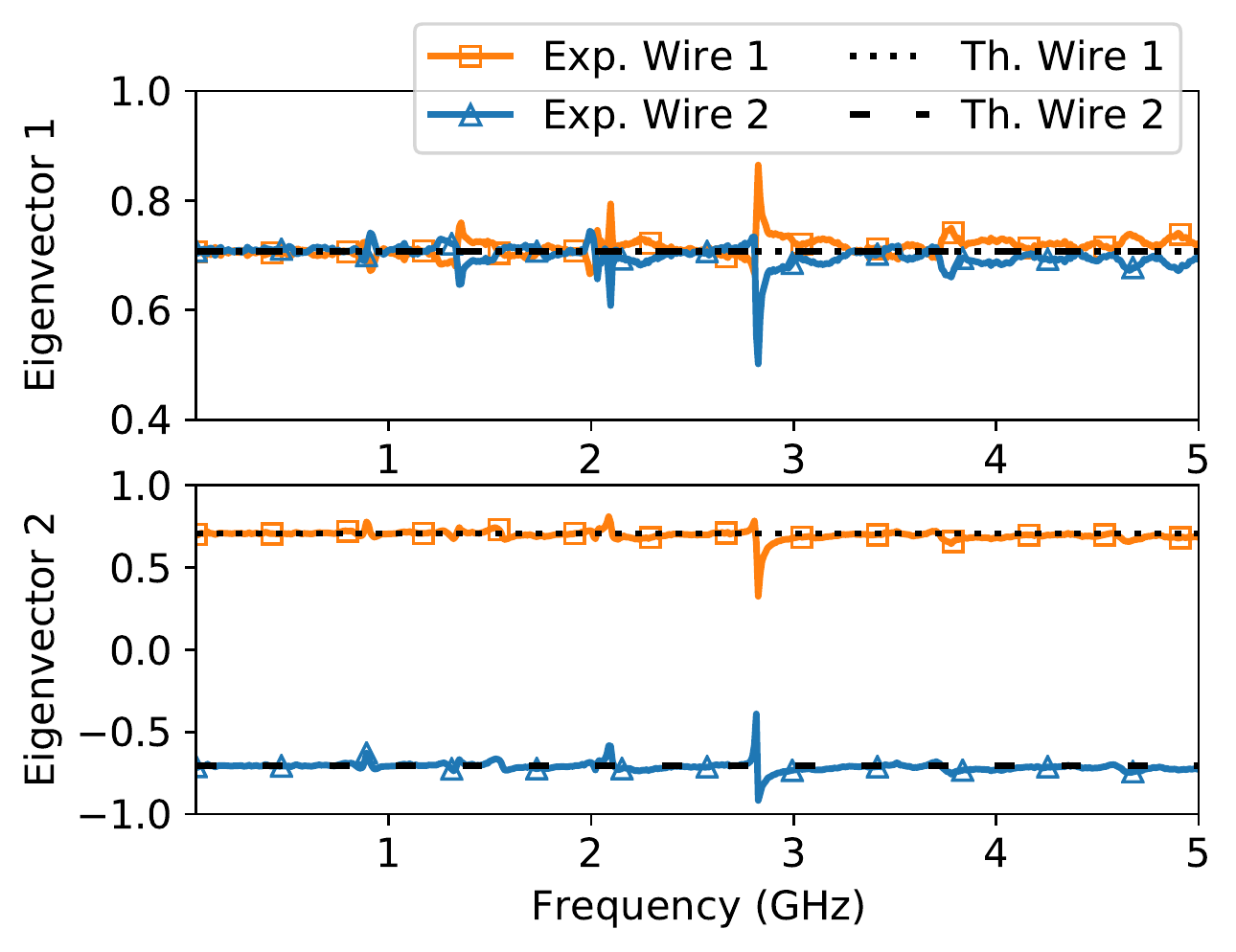}
		\caption{Normalised eigenvectors (real part) of the transfer matrix $\mathbf{H}$ of a two wire system obtained experimentally (Exp.) and theoretically (Th.). The eigenmodes represent a surface wave mode and differential mode, respectively.}
		\label{fig:Eigenmodes-2-Wires}
	\end{figure}
	
 	Having discussed the eigenmodes theoretically, we now provide experimental evidence supporting our model. To that end, Figure \ref{fig:Eigenmodes-2-Wires} shows the measured and calculated eigenmodes (real part) for two 0.5~mm annealed copper wires separated by $d=1$~cm. For the measurement, the eigenmodes were calculated using the MIMO approach discussed at the beginning of the section in the frequency range from 50~MHz to 5~GHz. Figure \ref{fig:Eigenmodes-2-Wires} shows good agreement between theory and experiment. Small ripples in the measured eigenvectors are attributed to resonances due to the finite length of the wires (50~cm) and discontinuities introduced by the launchers. Not shown is the relative phase between the conductors which except for a few peak values deviates less than $0.05\pi$ radians from its expected value of zero or $\pi$ for each mode respectively. Hence, the imaginary part of the measured eigenmodes is negligible. A plot of the imaginary part may be found in the Supplementary Material. 
	
	\begin{figure}[t]
		\centering
		\includegraphics[width=0.45\textwidth]{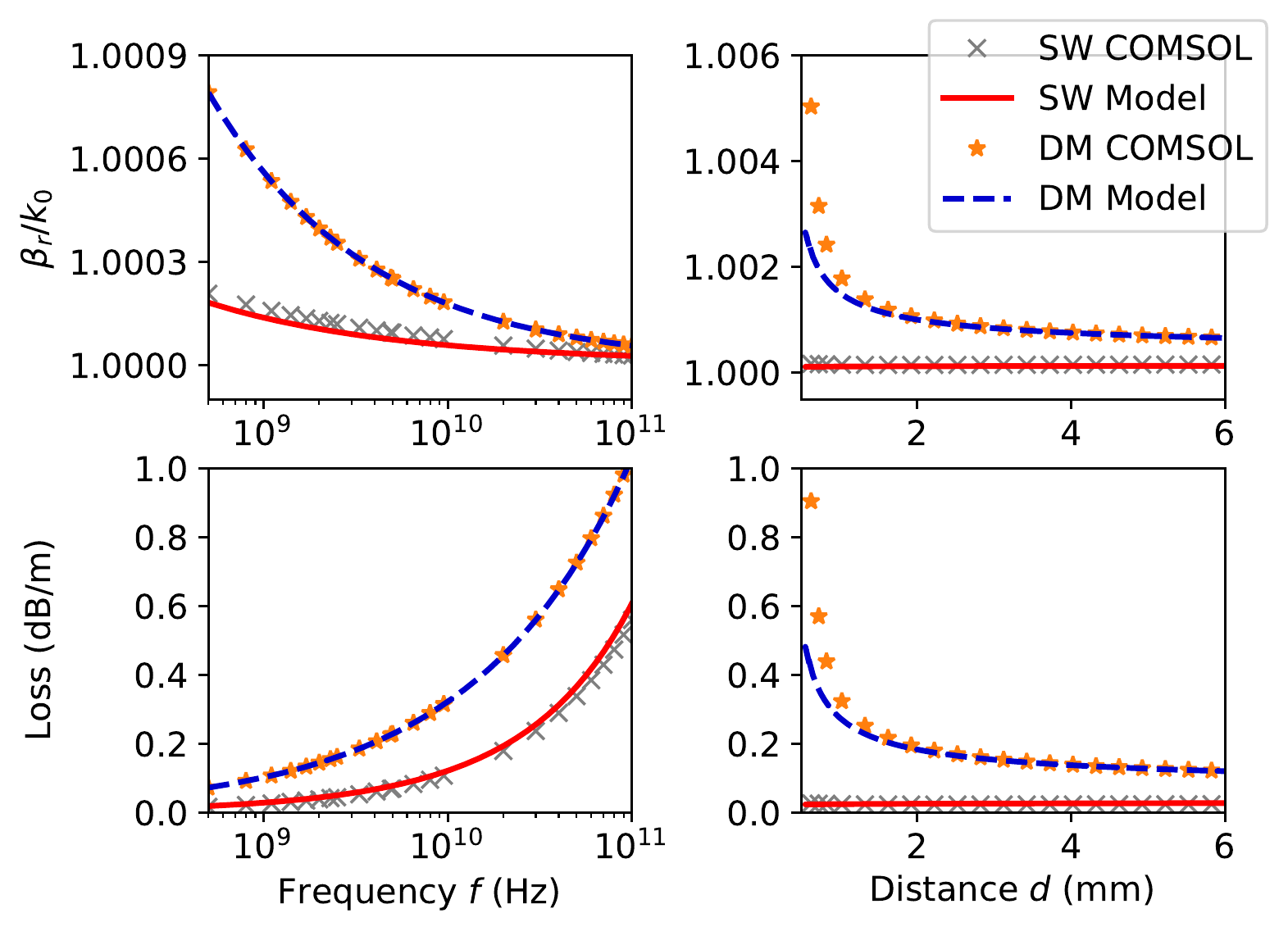}
		\caption{Phase constant and loss for a two wire line as a function of frequency ($d=1.05~\text{cm}$) and of distance ($f=1~\text{GHz}$) for the surface wave (SW) and differential mode (DM). Results were obtained using COMSOL and with the MTL model.}
		\label{fig:TwoWireSommerfeld}
	\end{figure}
	
	Next, we examined the propagation characteristics, namely loss and phase velocity, which for both modes are very similar in this case. Consequently, experimental measurement of these quantities is difficult especially because launcher de-embedding would need to be performed as well. Instead, we benchmark our model against the numerical solver COMSOL which uses the finite element technique. \cite{COMSOL-AB} This has the advantage that different parameters such as distance and frequency can be swept exactly and allows the determination of the complex propagation constant. The eigenmode solver in COMSOL finds the same eigenmodes, namely the differential mode and surface wave mode, as previously discussed. Thus, we can directly compare the values obtained with COMSOL to the values calculated using the MTL model. Details of the simulation can be found elsewhere.\cite{Molnar2021}
	
	The left hand side of Fig. \ref{fig:TwoWireSommerfeld} shows the phase constant and loss for the two conductor system as a function of frequency. The frequency is varied between $500~\text{MHz}$ and $100~\text{GHz}$ with the distance between the wires being fixed at $1.05~\text{cm}$. The results from COMSOL are very close to the predictions from the MTL model throughout this frequency range. 
	
	The right hand side of Fig. \ref{fig:TwoWireSommerfeld} shows the phase constant and loss as a function of distance for a frequency of $1~\text{GHz}$. For distances above 2~mm agreement is very good between the two methods. As the MTL model neglects proximity effects in contrast to COMSOL, the results deviate for smaller separations. However, the difference is a lot smaller for the surface wave than for the differential mode. Additionally, we can use these results to estimate applicability of the MTL model. For the given wire radius $a=0.25~\text{mm}$ results are accurate within a few percent for distances $d\geq 8a$.
	
	 We remark that for very high frequencies or large separations $(d\gg 1/\gamma)$, the two surface waves decouple and single wire surface waves emerge. Indeed, the MTL model reproduces these single wire solutions. Furthermore, when the system's lateral size becomes comparable to the wavelength, novel modes including radiating modes which are not acknowledged within our model may start propagating.\cite{Leviatan1982} However, the differential and surface wave mode will still be part of the complete mode spectrum. Hence, these results remain valid even at high frequencies although excitation of these modes individually may prove increasingly difficult.
	
	 Finally, we point out that, in general, the surface wave is seen to be significantly less lossy than the differential mode. While this is a promising avenue of application which has been exploited for instance by using surface waves as THz waveguides, it comes with the caveat that the surface wave easily radiates as discussed previously.\cite{Dfg2004, Jeon2005}
	
	\subsection{Three Wires}
	
	\begin{figure}
		\centering
		\includegraphics[width=0.45\textwidth]{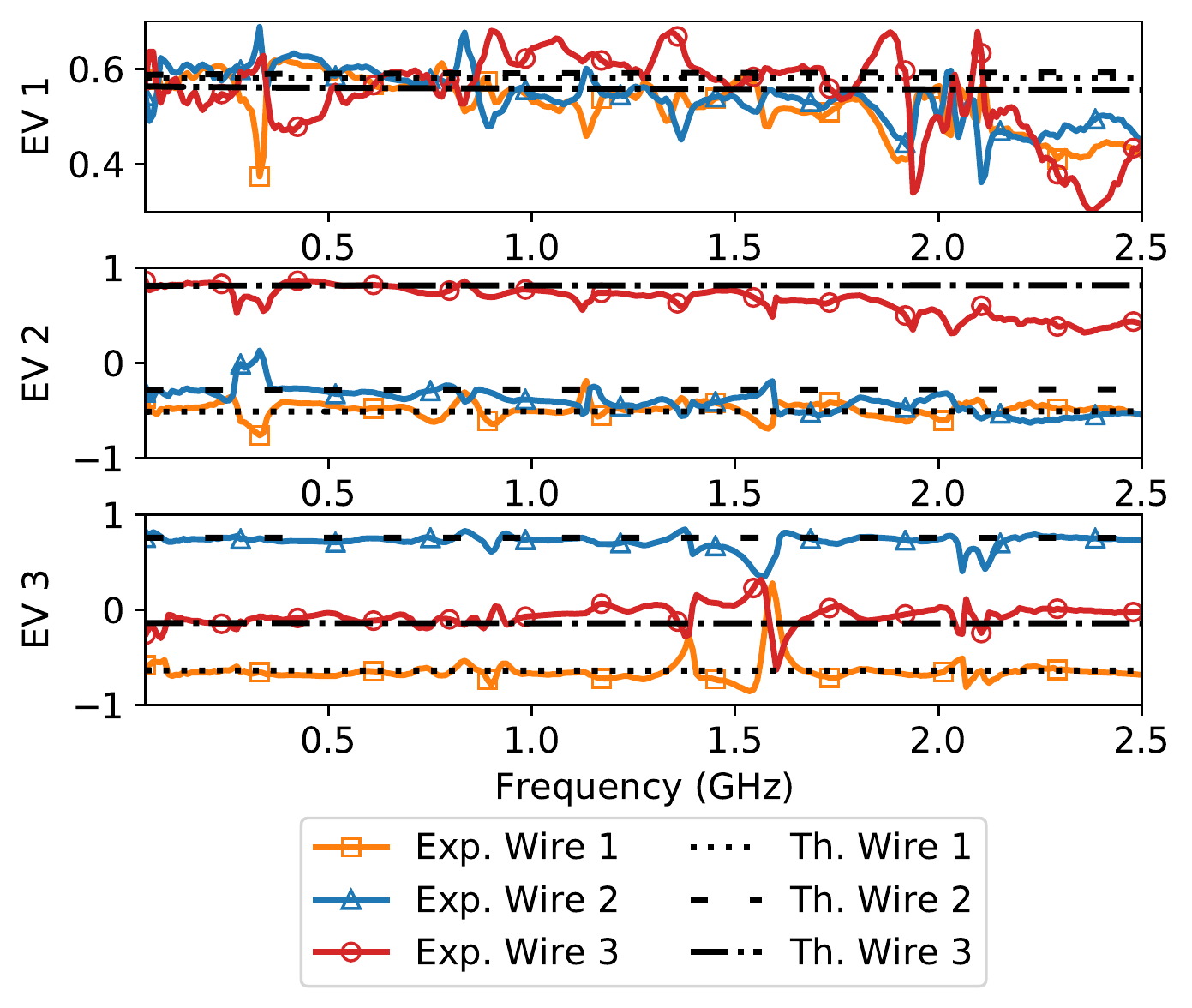}
		\caption{Real part of the three normalised eigenvectors (EV) of the transfer matrix $\mathbf{H}$ in a three wire system. Experimental (Exp.) and theoretical (Th.) results are shown.}
		\label{fig:Eigenmodes-3-Wire} 
	\end{figure}
	
	The three wire case is considerably more complex than the two wire case in that there is now no inherent symmetry in the per unit length parameter matrices. This has a benefit in that the extracted eigenmodes will depend on the exact geometry of the setup. In contrast, the eigenmodes of the two-wire case may have been predicted simply by symmetry. 
	
	In our experimental setup, there are three 0.5 mm copper wires aligned in a line. The outer two conductors are separated from the inner conductor by 1~cm and 2~cm, respectively. Using the MIMO decomposition described previously, we find the eigenmodes of this system experimentally in the range from 50~MHz to 2.5~GHz. Experimental and theoretical eigenvectors are shown in Fig. \ref{fig:Eigenmodes-3-Wire}. For frequencies below 1 GHz we see good, and up to 2 GHz reasonable agreement between theory and measurement. Above 2 GHz this agreement quickly deteriorates. Similar results are seen in the imaginary parts of the eigenvectors which are plotted in the Supplementary Material and generally increase with frequency. We attribute the increasing deviation to the large setup size (3~cm) which allows higher order modes to start propagating at lower frequencies.\cite{Leviatan1982} More specifically, above 1 GHz the lateral setup size is of the same order of magnitude as the wavelength. Ripples seen throughout the measurement are attributed in part to resonances caused by the launchers and the finite wire lengths. Furthermore, it proved difficult to maintain the nominal distance between the wires making the system non-uniform. These small changes in the relative wire distances can lead to mode conversion between the expected eigenmodes. 
	
	\section{Conclusion and Outlook}
	
	In this paper, we have outlined how surface waves can be incorporated into the framework of transmission line theory. Using the Sommerfeld surface wave, we defined voltage and current and then derived per unit length inductance, capacitance and impedance. Then, we extended this method to many conductors by assuming the fields may be approximated by a superposition of surface waves. It was shown that the currents and voltages are governed by the same telegrapher's equations as used in standard MTL theory. Accordingly, we used its solutions to calculate eigenmodes and propagation constants for two simple geometries: a 2-wire and a 3-wire setup. The eigenmodes were shown to agree with experimentally obtained eigenmodes up to frequencies where the lateral system size becomes comparable to the wavelength. Additionally, we showed that the derived propagation constant agrees well with numerical results further validating the theory. 
	
	Further studies in this field may include the study of coated wires and their surface wave solution namely the Goubau mode. Additional concrete examples of mode conversion in non-uniform cables and its effect on crosstalk could present another avenue of research. In particular, for the field of electromagnetic interference, surface waves which easily can be converted to radiation may pose a limiting factor for high frequency data transmission on the copper access network.
	
	In conclusion, this extended MTL theory can be used to simulate long cables up to MHz and even GHz frequencies. Thus, we hope it can provide some insights into the propagation characteristics of cables in the copper access network which may be used for novel high-frequency standards such as G.fast and G.mgfast. The additional information about the surface wave channel may hold significant value at high frequencies due to its smaller loss but increased radiation compared to the differential channel. 
	
	\section*{Supplementary Material}
	See Supplementary Material file for photos of the measurement setup and surface wave launchers as well as the imaginary parts of the eigenvectors from Figs. \ref{fig:Eigenmodes-2-Wires} and \ref{fig:Eigenmodes-3-Wire}.
	
%	\section*{Author's Contributions}
%	TS derived the theoretical model, conducted and analysed experiments and drafted the manuscript. DM carried out numerical modelling and revised the manuscript. AAR and MP supervised the research and critically revised the manuscript.
	
	% If you have acknowledgments, this puts in the proper section head.
    \begin{acknowledgments}
    This work was supported by the Royal Society Grants IF170002 and INF-PHD-180021. Additional funds were provided by BT plc and Huawei Technologies Düsseldorf GMBH. 
    \end{acknowledgments}
	
	\section*{Data Availability Statement}
	The code that support the findings of this study are available from the corresponding author upon reasonable request. The experimental data that support the findings of this study will be made openly available in Apollo at https://doi.org/10.17863/CAM.65674 upon acceptance. 
	
	\appendix
	
	\section{Approximate equality between $P_T$ and $P_{TL}$} \label{app:Pt=Ptl}
	Here we sketch a proof that $P_{TL}$ in \eqref{eq:PTL} approximates $P_T$ in \eqref{eq:PT} for good conductors where the wave travels close to the speed of light with low loss. We consider a conductor good if $\frac{\sigma}{\omega} \gg \epsilon_0$ such that $\epsilon_c$ can be approximated as $\epsilon_c \approx -i\frac{\sigma}{\omega}$. Furthermore, near lossless propagation close to the speed of light implies that $\gamma^2$ will be imaginary to leading order. So, it may be expressed as $\gamma^2\approx i \text{Im}(\gamma^2)$. Thus, we may approximate $P_{TL}$ as 
	\begin{equation}
		P_{TL}\approx\frac{1}{2}|B|^2\frac{2\pi a \epsilon_0}{|\gamma|^2} \text{Re}\Big(\omega \beta  \frac{\gamma H_0(\gamma a) H_1(\gamma a)^*}{i \text{Im}(\gamma^2)}\Big)
	\end{equation}
	As $\beta_r \gg \beta_i$, we may replace $\beta$ by its real part. Finally, using the identity $\text{Re}(-i x)=\text{Im}(x)$ for complex $x$, shows the approximate equality to $P_T$.
	
	\section{Calculation of Capacitance Matrix from Coefficients of Potential} \label{app:CfromP}
	The elements $C_{mn}$ define the capacitance matrix $\mathbf{C}$ and can be calculated from equation \eqref{eq:charge-MTL}. However, a simpler approach is calculating the coefficients of potential $P_{mn}$ which relate the per unit length charges to the voltage of conductor $m$ by
	\begin{equation}
		\sum_{n=1}^N P_{mn} \lim\limits_{\Delta z \rightarrow 0}\big(\frac{Q_n}{\Delta z}\big) = V_m. \label{eq:P-definition}
	\end{equation}
	The matrix $\mathbf{P}$ consisting of matrix elements $P_{mn}$ is the inverse of the capacitance matrix $\mathbf{C}$, i.e. $\mathbf{C}=\mathbf{P}^{-1}$ (Ref. \onlinecite{Paul1994}). Thus, assuming $\mathbf{P}$ is invertible, calculating $\mathbf{P}$ is equivalent to determining $\mathbf{C}$. Any element $P_{mn}$ can be calculated by forcing all charges except $Q_n$ to zero in \eqref{eq:P-definition}. As the charges cause surface waves, only a surface wave on conductor $n$ contributes to the voltage of conductor $m$. For $n=m$ the voltage is simply the single wire voltage \eqref{eq:voltage}. For $n\neq m$ the voltage is
	\begin{equation}
		V_m|_{Q_1,...,Q_N=0, Q_n\neq0}=\int_{d_{mn}}^{\infty}\vec{E}_n \cdot \hat{r} dr
	\end{equation}
	with radial unit vector $\hat{r}$ and where $\vec{E}_n$ is the electric field of a surface wave on conductor $n$.	Using \eqref{eq:charge} for the per unit length charge of a single surface wave, we find the following expressions:
	\begin{equation}
		P_{mn}=\begin{cases}
			\frac{H_0(\gamma d_{mn})}{2\pi \epsilon_0  \gamma a H_1(\gamma a) (1-\frac{\epsilon_0}{\epsilon_c}))} \quad \text{for }m\neq n\\
			\frac{H_0(\gamma a)}{2\pi \epsilon_0 \gamma a  H_1(\gamma a) (1-\frac{\epsilon_0}{\epsilon_c}))} \quad \text{for }m= n
		\end{cases}
	\end{equation}
	Finally, we note that for good conductors at frequencies below a few THz, $|\epsilon_c|\approx \frac{\sigma}{\omega}   \gg \epsilon_0$ in which case $\mathbf{P}\approx \frac{1}{\mu_0 \epsilon_0} \mathbf{L}$ and hence $\mathbf{C} \approx \mu_0 \epsilon_0 \mathbf{L}^{-1}$
	
% Create the reference section using BibTeX:

\bibliographystyle{aipnum4-1}
%\bibliography{library}

%merlin.mbs aipnum4-1.bst 2010-07-25 4.21a (PWD, AO, DPC) hacked
%Control: key (0)
%Control: author (8) initials jnrlst
%Control: editor formatted (1) identically to author
%Control: production of article title (-1) disabled
%Control: page (0) single
%Control: year (1) truncated
%Control: production of eprint (0) enabled
%

\end{document}